\documentclass[useAMS,usenatbib,fleqn]{mnras}
\usepackage[utf8]{inputenc}
\usepackage{graphicx,color}
\usepackage{mathtools}	
\usepackage{amsmath}
\usepackage{amssymb}
\usepackage{bm}
\usepackage{pstricks} 
\usepackage{arydshln} 
\usepackage[normalem]{ulem}
    \usepackage[british]{babel}             
    \usepackage[slantedGreek]{newtxmath}    
    \usepackage[T1]{fontenc}                

\def\apj{ApJ}

\def\mnras{MNRAS}
\def\aap{A\&A}
\def\apjl{ApJ}

\def\apjs{ApJS}

\def\nat{Nature}
\def\pasp{PASP}

\def\aj{AJ}

\def\aapr{ARA\&A}

\newcommand{\Edd}{_\mathrm{Edd}}

\newcommand{\Rsun}{\,\mathrm{R_\odot}}
\newcommand{\Msun}{\,\mathrm{M_\odot}}

\newcommand{\rmD}{\mathrm{\Delta}}

\newcommand{\MJ}{\,\mathrm{M_\mathrm{J}}}

\newcommand{\w}{_\mathrm{wd}}

\newcommand{\eone}{_\mathrm{e,1}}
\newcommand{\etwo}{_\mathrm{e,2}}

\newcommand{\ione}{_\mathrm{i,1}}
\newcommand{\itwo}{_\mathrm{i,2}}
\newcommand{\fone}{_\mathrm{f,1}}
\newcommand{\ftwo}{_\mathrm{f,2}}
\newcommand{\done}{_\mathrm{d,1}}

\newcommand{\ej}{_\mathrm{ej}}
\newcommand{\thm}{_\mathrm{thm}}
\newcommand{\inj}{_\mathrm{inj}}
\newcommand{\rad}{_\mathrm{rad}}

\newcommand{\esctwo}{_\mathrm{esc,2}}
\newcommand{\twomax}{_\mathrm{2,max}}
\newcommand{\twothm}{_\mathrm{2,thm}}

  \newcommand{\K}{\,{\rm K}}

  \newcommand{\Myr}{\,{\rm Myr}}

  \newcommand{\yr}{\,{\rm yr}}     
  \newcommand{\da}{\,{\rm d}}

\title[Successive common envelope events] 
{Successive common envelope events from multiple planets}
\author[L.~Chamandy et al.]{Luke Chamandy,$^1$\thanks{lchamandy@pas.rochester.edu}
Eric G.~Blackman,$^1$\thanks{blackman@pas.rochester.edu}
Jason Nordhaus$^{2,3}$\thanks{nordhaus@astro.rit.edu}
and Emily Wilson$^2$\thanks{ecw7497@rit.edu}
\\
$^1$Department of Physics and Astronomy, University of Rochester, Rochester NY 14627, USA\\
$^2$Center for Computational Relativity and Gravitation, Rochester Institute of Technology, Rochester, NY 14623, USA\\
$^3$National Technical Institute for the Deaf, Rochester Institute of Technology, Rochester, NY 14623, USA
}
\begin{document}
\maketitle
\begin{abstract}
Many stars harbour multi-planet systems. As these stars expand late in their evolutions, the innermost planet may be engulfed, leading to a common envelope (CE) event. Even if this is insufficient to eject the envelope, it may expand the star further, causing additional CE events, with the last one unbinding what remains of the envelope. This multi-planet CE scenario may have broad implications for stellar and planetary evolution across a range of systems.  We develop a simplified version and show that it may be able to explain the recently observed planet WD~1856~b.
\end{abstract}
\defcitealias{Vanderburg+20}{V20}
\defcitealias{Chamandy+18}{Paper~I}
\defcitealias{Chamandy+19a}{Paper~II}
\defcitealias{Chamandy+19b}{Paper~III}
%
\section{Introduction}
\label{sec:intro}
\citet{Vanderburg+20} (hereafter~\citetalias{Vanderburg+20}) recently detected the
Jupiter-sized object WD~1856+534~b (hereafter WD~1856~b) orbiting a white dwarf (WD) with orbital period  $1.4\da$.  
They calculated the WD mass to be $(0.518\pm0.055) \Msun$,
and obtained an  upper limit of $13.8\MJ$ implying that it is likely a planet.
The semi-major axis of $\sim4\Rsun$ places it well within the envelope of the progenitor star, 
but inside of the WD-planet period gap predicted by \citet{Nordhaus+10}.%
\footnote{The WD-planet period gap is the orbital region around a WD 
that is predicted to be devoid of planets due to previous interactions during stellar evolution.}
A single common envelope (CE) event origin is thus unlikely \citep{Nordhaus+Spiegel13}.    
Likewise, \citetalias{Vanderburg+20} 
found that a single CE origin is unlikely. 
\citet{Lagos+20} argued that a single CE origin is not ruled out 
if energy additional to the released orbital energy contributes to envelope unbinding. 
They also argue that a CE scenario is consistent with the youth 
needed to account for apparent membership in the Galactic thin disc.

Using Modules for Experiments in Stellar Astrophysics (MESA) \citep{Paxton+19}, 
we also assessed the single CE scenario based on detailed stellar interior models 
which match the initial-final mass relation \citep{Cummings+18}.
Assuming that orbital energy alone is used to eject the envelope with maximum efficiency 
\citep{Wilson+Nordhaus19,Wilson+Nordhaus20} we 
find that the stellar envelope around a WD core of mass $(0.518 \pm 0.055) \Msun$ 
can be ejected by a companion of $13.8\MJ$ or less only if the initial primary mass is $<1.2\Msun$.
Successful single CE scenarios 
resulting in the observed $1.4\da$ period can only occur 
during the late stages of the asymptotic giant branch (AGB) phase
with $<0.1\Msun$ left in the envelope and average envelope-mass-to-core-mass ratio $<0.14$.
But for this stellar mass range, the CE would   initiate on the red giant branch (RGB), 
leaving a WD of too little mass. This system is thus unlikely the result of a single CE interaction.

\citetalias{Vanderburg+20} consider ideas other than CE to explain the orbit of WD~1856~b,
and favour planet migration from a  larger orbit by the von Zeipel-Lidov-Kozai (ZLK) effect
\citep{vonZeipel1910,Lidov62,Kozai62} in tandem with tidal friction \citep{Wu+Murray03,Fabrycky+Tremaine07}.
They conclude that  WD~1856~b's M-dwarf companions G~229-20~A/B were unlikely to
 have triggered the migration, but  that  unseen planets could have done so.
By contrast,  \citet{Munoz+Petrovich20}, \citet{Oconnor+21} and \citet{Stephan+20} 
find regions of parameter space for which the M-dwarf companions can excite the ZLK effect, 
but disagree on the importance of various terms and the initial semi-major axis of WD~1856~b.

WD~J0914 may also  host a planet \citep{Gansicke+19}, 
as inferred by modeling observational signatures of accretion.
The cooling age of WD~J0914 is only $\sim13\Myr$, 
leaving little time for migration to its present orbit of semi-major axis $\sim15\Rsun$.
The ZLK-tidal friction mechanism is a possible explanation, 
but it may require the presence of an unseen companion with mass $\gtrsim0.3\Msun$ \citep{Stephan+20}.
A single CE origin may  be possible if the planet  mass $\gtrsim1\MJ$,
but the mass may  be smaller \citep{Gansicke+19}.

Here we devise a model that involves successive planet-induced CE events to explain WD~1856~b.
Such a scenario was mentioned by \citet{Nordhaus+10}, 
and \citet{Lagos+20} in the context of WD~1856~b,  but has not yet been quantified. 
Here we demonstrate  that such a model is efficacious. 
We focus on  WD~1856, but the model can be applied to other systems  such as WD~J0914.
The scenario starts with a CE interaction involving planet~1 -- 
which is subsequently tidally disrupted -- and the post-main sequence WD progenitor.
Different pathways to the first CE (CE1) \citep[e.g.][]{Villaver+Livio09,Chen+18,Macleod+18a} apply to planet~1, 
but subsequent rapid expansion of the primary engulfs the wider orbit of planet-2, causing a second CE (CE2). 
We assume that  CE1 and CE2 combine to remove the envelope,
and planet~2 survives in orbit.\footnote{The scenario can be  generalized for more planets.}

The  observed mass of WD~1856  constrains the zero-age main sequence mass of its progenitor, $M_0$.
For $M_0\lesssim1.0\Msun$, the maximum RGB radius exceeds that of the AGB,
so CE will occur on the RGB, and leave too small of a  WD core. For $M_0\gtrsim3.5\Msun$, 
the core is too massive on the AGB, but, coincidentally, attains high enough mass on the RGB.
But a high mass star, higher envelope binding energy, and smaller RGB vs. AGB radius make CE in this mass range less likely. 
For $1.0\lesssim M_0/\!\Msun\lesssim3.5$, CE  likely occurs on the AGB with a core mass $M\w=(0.518\pm0.055)\Msun$.
We expect CE to occur before the AGB tip. There, a $1\Msun$ star has lost $\sim0.4\Msun$,
and a $3.5\Msun$ star has lost little before the start of the AGB.  
Therefore, the AGB mass before CE1 is $0.6\lesssim M_1/\!\Msun\lesssim3.5$.

In Sec.~\ref{sec:model}, 
we compute the respective fractions of the initial envelope binding energy injected during CE2 and CE1 
for a given mass $m_2$ of planet~2.
We then determine the mass $m_1$ of planet~1, for limiting cases to bracket its range.
For plausible models, CE1 should cause CE2.
We estimate the expansion of the star during CE1,
and show that it can be large enough for the star to engulf the second, more distant planet~2.
In Sec.~\ref{sec:cooling}, we incorporate radiative cooling.

\vspace{-0.5cm}
\section{Two-planet CE scenario}
\label{sec:model}
\subsection{Planet mass combinations for two limiting cases}
\label{sec:mass}
To exemplify, we use the WD mass, radius, and planet~2 orbital radius of \citetalias{Vanderburg+20}: 
$M\w=0.518\Msun$, $R\w=1.31\times10^{-2}\Rsun$, and $P_2= 1.408\da$. 
Subscript `1' and '2' will delineate  the state of the system at the start of CE1 and CE2 respectively.
Each planet mass is assumed  constant, but we allow  tidal disruption of planet~1.
The binding energy (defined $>0$) of the envelope at the start of mass transfer that leads to CE1 is 
\begin{equation}
  \label{E1}
  E_1= \frac{GM_1 M\eone}{\lambda_1R_1},
\end{equation}
where $G$ is the gravitational constant,
$M_1$ and $M\eone$ are  the original masses of the primary and  envelope; 
$R_1$ is the original radius of the primary; and 
$\lambda_1$ is a conventional dimensionless parameter accounting for distinct radial energy profiles. 
If a fraction $\beta_1$ of the energy needed to unbind the envelope is supplied by CE1,
then CE2 need supply only a fraction $\beta_2=1-\beta_1$ to unbind the remaining envelope.
The CE energy formalism \citep[e.g][]{Ivanova+13a} applied to CE1 then gives
\begin{equation}
  \label{CE1}
  \beta_1 E_1= (1-\beta_2)E_1\simeq \alpha_1 \frac{G M\w m_1}{2a\fone},
\end{equation}
where $\alpha_1$ is a conventional dimensionless parameter accounting for the energy conversion efficiency of the unbinding,
$m_1$ is the mass of planet~1, the primary core mass is assumed equal to $M\w$, 
$a\fone$ is the binary separation at the end of CE1,
the small mass of the envelope interior to the orbit has been neglected (or can be absorbed into the factor $\beta_1$),
and the term $-\alpha_1 G M_1 m_1/2a\ione$ on the right side (with $a\ione$ the initial separation) is neglected as in \citetalias{Vanderburg+20}; 
we have checked that this term is generally small.
Similarly, for CE2
\begin{equation}
  \label{CE2}
  \beta_2 E_1\simeq \alpha_2\frac{GM\w m_2}{2a\ftwo}.
\end{equation}
Dividing equation~\eqref{CE1} by equation~\eqref{CE2}, and rearranging gives
\begin{equation}
  \label{mcomprime}
  m_1= \frac{a\fone}{a\ftwo}\frac{\alpha_2}{\alpha_1}\left(\frac{1}{\beta_2}-1\right)m_2.
\end{equation}
\citetalias{Vanderburg+20} constrained $m_2\le 14\MJ$, though it is not  precisely known.
Dividing equation~\eqref{CE2} by equation~\eqref{E1}  gives
\begin{equation}
  \label{beta}
  \beta_2= \frac{M\w m_2}{2(M_1-M\w)M_1}\frac{\alpha_1\lambda_1R_1}{a\ftwo}\frac{\alpha_2}{\alpha_1},
\end{equation}
where we have substituted $M\eone=M_1-M\w$.
The quantity $a\ftwo$ is related to the observed period by Kepler's third law:
\begin{equation}
  \label{af}
  a\ftwo= \left[G(M\w+m_2)\left(\frac{P\ftwo}{2\pi}\right)^2\right]^{1/3} \simeq \left(\frac{GM\w P\ftwo^2}{4\pi^2}\right)^{1/3},
\end{equation}
where $P\ftwo=1.4\da$ is the observed period.
In the rightmost expression and  below, we  assume $m_1,m_2 \ll M\w$.

\begin{figure}
  \includegraphics[width=0.475\textwidth,clip=true,trim= 60 115 0 45]{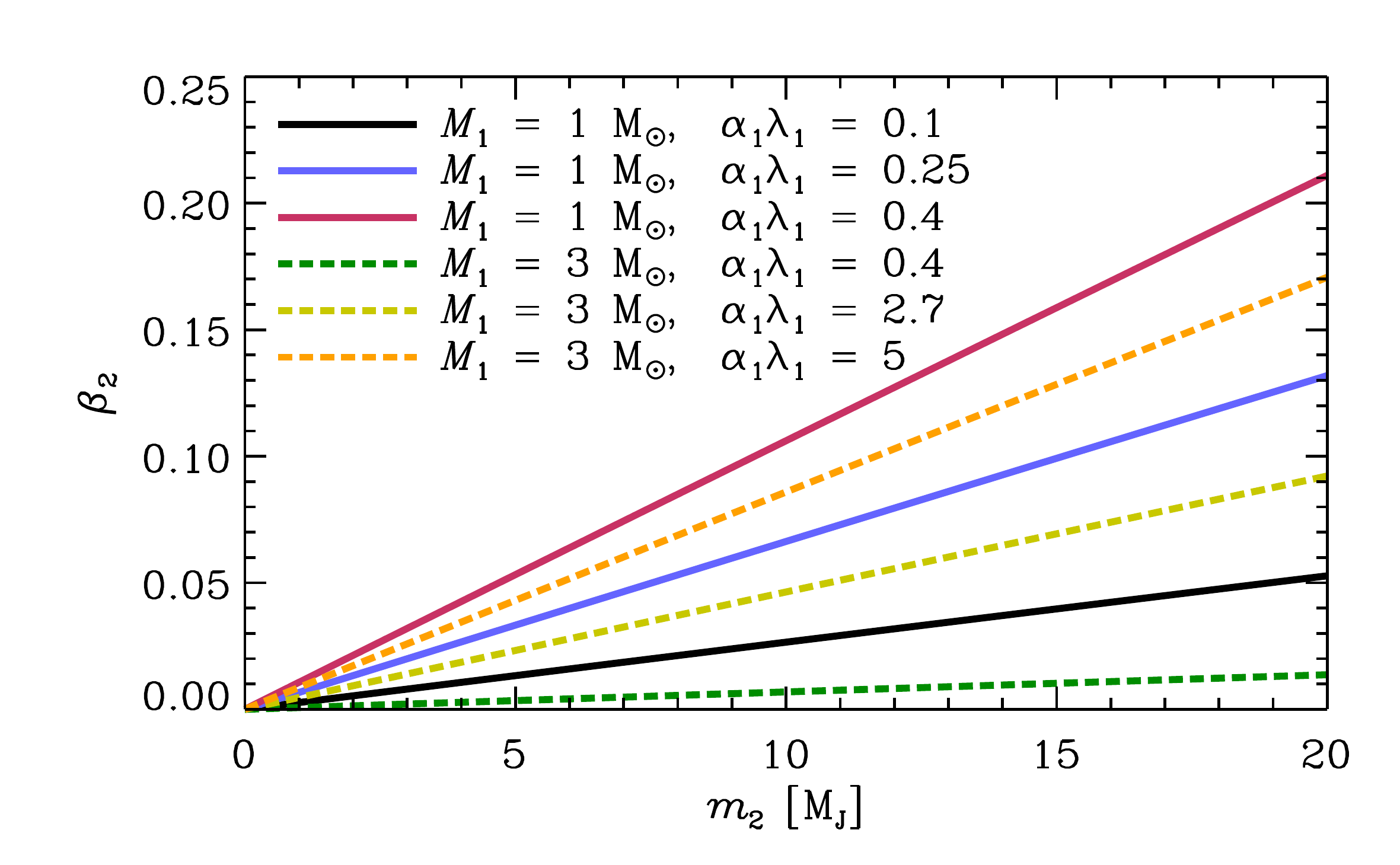}\\
  \includegraphics[width=0.475\textwidth,clip=true,trim= 60  20 0 45]{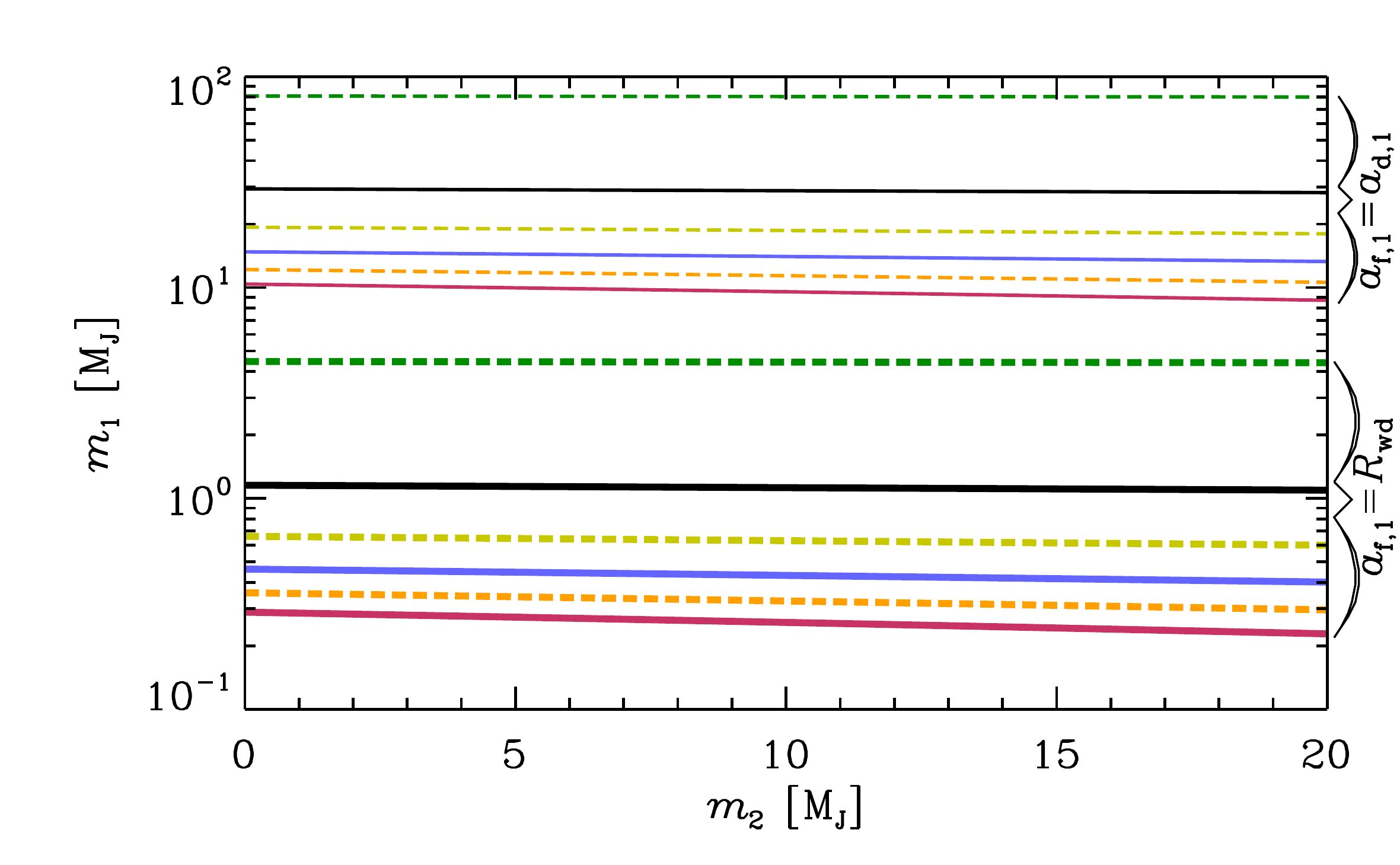}\\
  \vspace{-0.5cm}
  \caption{\textit{Top:}~The fraction $\beta_2$ of the original binding energy $E_1$ injected during CE2,
           versus the mass of planet~2. 
           \textit{Bottom:}~The mass $m_1$ of planet~1 required to inject the fraction $\beta_1=1-\beta_2$ 
           of the binding energy $E_1$,
           assuming that the final separation is equal to the WD radius, $a\fone=R\w$ (thick lines), 
           or to the tidal disruption separation, $a\fone=a\done$ (thin lines).
           The latter is calculated assuming that the planet radius $r_1=0.1\Rsun$.
           The lines are not exactly horizontal.
           In both panels $\alpha_1/\alpha_2=1$ is used.
           \citetalias{Vanderburg+20} constrain the mass of WD~1856~b to be  $0.1\lesssim m_2/\!\MJ\lesssim13.8$.
           \label{fig:alp_12_ratio_1}
          }            
          \vspace{-0.3cm}
\end{figure}

In the top panel of Fig.~\ref{fig:alp_12_ratio_1} we plot $\beta_2$ against $m_2$, 
in Jupiter masses, for various values of $\alpha_1\lambda_1$
chosen to be consistent with estimates of \citetalias{Vanderburg+20}, for either $M_1=1\Msun$ or $M_1=3\Msun$.
Since $\beta_2<1$ for $m_2<14\MJ$, a single CE  scenario fails, in agreement with \citetalias{Vanderburg+20}. 
On the other hand, extremely small values of $\beta_2$ would require fine-tuning $\beta_1$ to be just less than $1$.
We see that $\beta_2$ can take on values up to $0.15$ for $m_2<14\Msun$.
Focusing on $\beta_2 >0.01$ gives $\alpha_1\lambda_1\gtrsim0.4$ in the $3\Msun$ model 
and also determines a corresponding lower limit on $m_2$.

Following \citetalias{Vanderburg+20}, 
we assume  $R_1$ to be equal to the primary's Roche lobe when mass transfer initiates, 
and then make use of their fitting formula for $R_1$ in terms of $M\w$.
This gives
\begin{equation}
  \label{Rp}
  R_1= 5.56\times10^4 f(M\w/\!\Msun) \Rsun,
\end{equation}
where $f(\mu)\equiv \mu^{19/3}/(1+20\mu^3+10\mu^6)+f_0$, 
with $f_0= 7.2\times10^{-5}$.

For $a\fone$, we consider two limiting cases: 
(i) the planet merges with the WD, with all of the liberated orbital energy released to the envelope so
\begin{equation}
  \label{af1casei}
  a\fone\simeq R\w \approx 1.3\times10^{-2}\Rsun \qquad \mathrm{(case\:i)},
\end{equation}
and (ii) after the planet gets tidally disrupted, its orbital energy is no longer used to unbind the envelope. 
In this case, we can estimate \citep[c.f.][]{Nordhaus+Blackman06}
\begin{equation}
  \begin{split}
  \label{af1caseii}
  a\fone&= a\done\simeq \left(\frac{2M\w}{m_1}\right)^{1/3}r_1 \qquad \mathrm{(case\:ii)}\\
  &\simeq1.0\Rsun \left(\frac{M\w}{0.52\Msun}\right)^{1/3}
  \left(\frac{m_1}{10^{-3}\Msun}\right)^{-1/3}
  \left(\frac{r_1}{0.1\Rsun}\right),
  \end{split}
\end{equation}
where $a\done$ is the tidal disruption radius, measured from the centre of the primary's core, and $r_1$ is the radius of planet~1.
Cases (i) and (ii) bracket the range of possibilities.

Equation~\eqref{mcomprime}, with equations~\eqref{beta}, \eqref{af}, \eqref{Rp}, and \eqref{af1casei} or \eqref{af1caseii},
are then used to obtain $m_1$ in terms of $\alpha_1\lambda_1$, $a\fone$, $m_2$, $P\ftwo$, $M\w$, $M_1$ and the ratio $\alpha_2/\alpha_1$.
The results are  plotted in the bottom panel of Fig.~\ref{fig:alp_12_ratio_1}, 
for $\alpha_1/\alpha_2=1$ and the parameter values mentioned in the figure legend and caption.
Case~(i), $a\fone=R\w$, is represented by thick lines, and thin lines show case~(ii), $a\fone=a\done$, 
assuming $r_1=0.1\Rsun$. 
For $M_1=1\Msun$, we see that the required planet mass 
needed to inject the fraction $\beta_1$ of the original envelope binding energy 
falls within a reasonable range of $0.2\lesssim m_1/\MJ \lesssim30$.
For $M_1=3\Msun$ we obtain $0.3\lesssim m_1/\MJ\lesssim5$ if $a\fone=R\w$, 
but $10\lesssim m_1/\MJ\lesssim80$ if $a\fone=a\done$ is adopted. 
The latter range falls mostly above the usual planet-brown dwarf boundary of $\sim13\MJ$.

Fig.~\ref{fig:alp_12_ratio_1} shows that there is a large section of parameter space 
that produces $\beta_2>0.01$ and realistic values for $m_1$, viz.~$0.2\lesssim m_1/\MJ\lesssim30$.
The smaller value of $M_1=1\Msun$ (solid lines) leads to a larger viable parameter space than with $M_1=3\Msun$ 
(dashed lines).
The $1\Msun$ value may be favoured 
because the initial mass function is weighted toward lower-mass progenitors \citep{Chabrier03}.

\vspace{-0.3cm}
\subsection{Expansion of the envelope during CE1}
\label{sec:expansion_nocooling}
While it is possible that the primary star could expand on its canonical evolutionary time ($\gtrsim\!\Myr$)
to eventually engulf planet~2, 
this time is much longer than typical CE plunge times of days to years \citep[e.g.][]{Chamandy+20}.
Here we explain why  planet~2 could be engulfed as a \textit{result} of CE1.

During CE1, energy is predominantly deposited at the base of the envelope, 
due to the $1/r$ potential and centrally condensed evolved star.
We expect the envelope to  respond  by expanding.
To estimate the  expansion, 
we equate the initial energy just after CE1 with the final energy after the envelope has adjusted. 
The envelope might oscillate \citep{Clayton+17}, 
but we are only interested in the maximum radius reached.
We also neglect any recombination energy release during expansion. 
We do not include a bulk kinetic energy term for the bound envelope but allow for escaping winds.
We also do not consider the consequences of  accretion for nuclear fusion  \citep{Siess+Livio99a,Siess+Livio99b}.\footnote{Fusion releasing a fraction $f$ of rest energy can energize more than accretion if $\gtrsim {GM\w/(fc^2R\w)}$ of accreted  mass fuses.}
We obtain
\begin{equation}
  \label{E}
  -(1-\beta_1)\frac{GM_1 M\eone}{\lambda_1R_1}= -\frac{GM_2 M\etwo}{\lambda_2R_2}+E\ej.
\end{equation}
where $E\ej$ is the energy of the ejecta (wind), which is not well constrained.
If the ejecta leaves at the escape speed, it will remove only a net thermal energy, 
as bulk kinetic and potential energies sum to zero.
For convenience, we write 
\begin{equation}
  \label{Eej}
  E\ej= \frac{1}{2}C' \frac{M\ej v\esctwo^2}{\lambda_2} =\frac{C' GM_2 (M\eone-M\etwo)}{\lambda_2R_2},
\end{equation}
with $M\ej=M\eone-M\etwo$ the ejecta mass, $v\esctwo$ the escape speed from the stellar surface  in its final state, 
and $C'$ a dimensionless constant of order unity.
Substituting equation~\eqref{Eej} into equation~\eqref{E},
and defining $\lambda_2'\equiv \lambda_2 M\etwo/(M\etwo-C'M\ej)$, 
the right  of equation~\eqref{E} becomes $-GM_2 M\etwo/\lambda_2' R_2$.
Note that $\lambda_2'\longrightarrow\lambda_2$ as $C'$ or $E\ej\longrightarrow0$.
Rearranging equation~\eqref{E} gives
\begin{equation}
  \label{r2}
  \frac{R_2}{R_1}=\frac{1}{\beta_2}\frac{\lambda_1}{\lambda_2'}\frac{M_2}{M_1}\frac{M\etwo}{M\eone}.
\end{equation}
The mass of the ejecta $M\ej$ cannot exceed $M\eone/(1+C')$ or else $R_2$ reduces to zero;
we focus on solutions for which $R_2>R_1$, which requires $M\ej$ to be less than some smaller critical value.
If $M\ej=0$, then $M\etwo=M\eone$ and the solution reduces to
\begin{equation}
  \label{Rratio}
  \frac{R_2}{R_1}=\frac{1}{\beta_2}\frac{\lambda_1}{\lambda_2}; \qquad \mathrm{(no\:mass\:loss)}.
\end{equation}

As example~I, we take $M_1=1\Msun$ so that $M\eone=M_1-M\w\approx0.48\Msun$, 
and adopt $\lambda_1=\lambda_2$, $C'=1$, and $\beta_2=0.1$.
Assuming no mass loss, we obtain $R_2=10R_1$.
The maximum ejecta mass ($R_2=0$) is $\approx0.24\Msun$, whilst $R_2=R_1$ for $M\ej\approx0.21\Msun$.
If $M\ej=0.1\Msun$, we obtain $R_2=5.3R_1$.
As example~II, we take $M_1=3\Msun$, which gives $M\eone\approx2.48\Msun$,  
and  choose $\lambda_2=3\lambda_1$ \citep[c.f.][]{Xu+Li10}, $C'=3$,
and $M\ej=0.3\Msun$. 
Then we obtain $R_2=1.5R_1$.
For $M\ej=0$, we obtain $R_2=3.3R_1$.
This exemplifies cases where the envelope expands enough to engulf a second planet.

The maximum expansion of the primary during CE1 
constrains the maximum initial orbital radius of planet~2 to be engulfed.
For small orbital eccentricities and stellar mass loss,
mutual planet-planet perturbations are not expected 
to greatly restrict the mutual proximity of the planets \citep{Hill1886,Debes+Sigurdsson02,Maldonado+20}.
However, assuming that $a\itwo>a\ione$, CE1 requires  $a\ione<R_2$.
\citet{Villaver+Livio09} determine the \textit{maximum} $a\ione$ that still allows CE during the RGB 
for a Jupiter-mass planet with different primary masses (neglecting  other planet influences).
For $M_1=1\Msun$ and $3\Msun$, they obtain  $\max(a\ione)/R_1\approx2.5$ and $1.2$, respectively.
Values of $R_2/R_1$ from our model  comfortably exceed these requirements.
We do not include increases in $a\itwo$ from mass loss \citep[e.g.][]{Veras+11} and decreases associated  
from consequent drag on planet~2.

\vspace{-0.4cm}
\section{Role of cooling}
\label{sec:cooling}

\subsection{Expansion including cooling}
\label{sec:expansion_withcooling}
Not all of the energy injected into the envelope during the CE phase contributes  to unbinding it.  
This includes, but is not limited to, energy lost via radiation that would have otherwise assisted unbinding.
This is among the inefficiencies   included  in the $\alpha$-parameter of equations~\eqref{CE1} and \eqref{CE2}.

However, if cooling is as fast as energy injection, 
a ``self-regulated'' state could arise with the energy injected swiftly radiated 
\citep{Meyer+Meyer-hofmeister79,Ivanova+13a},
and the expansion of CE1 quenched. 
Here we modify the calculation of Sec.~\ref{sec:expansion_nocooling} to account for this.

\begin{figure}
  \includegraphics[width=0.48\textwidth,clip=true,trim= 40   0 0 45]{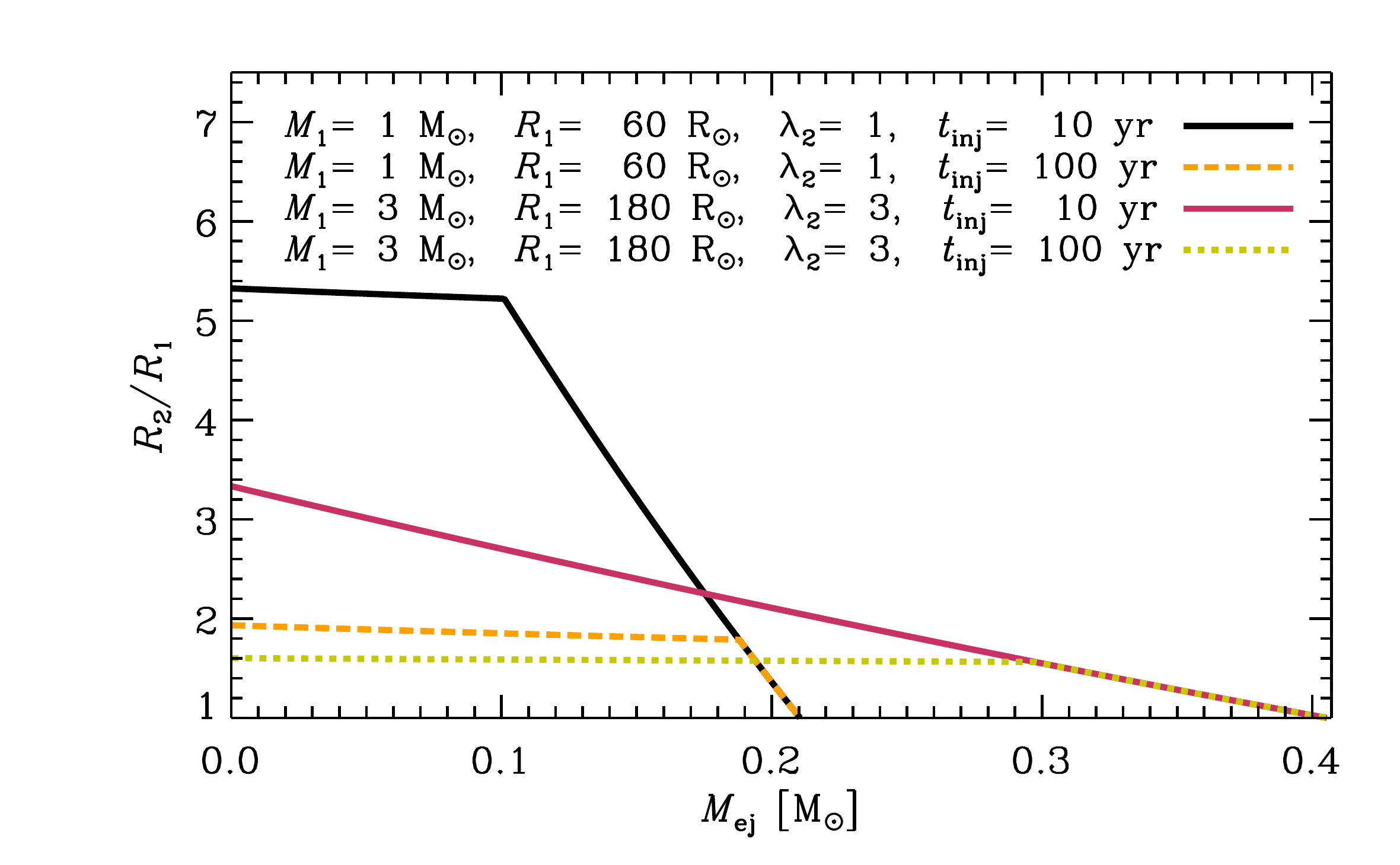}\\
  \includegraphics[width=0.48\textwidth,clip=true,trim= 40  20 0 45]{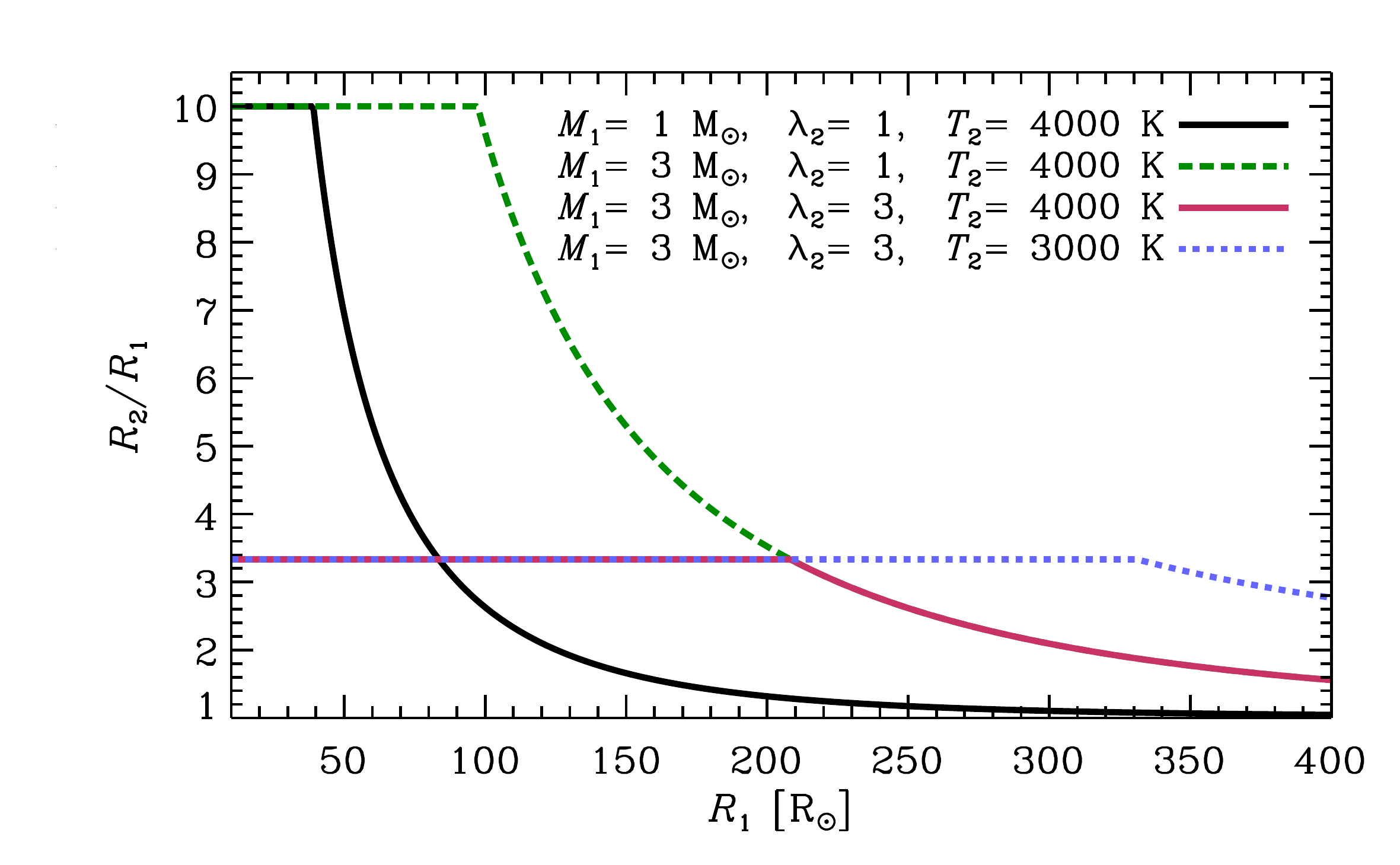}	
  \caption{\textit{Top}:~The ratio $R_2/R_1$ of the stellar radius before and after CE1 (which occurs on the AGB), 
           as a function of the ejecta mass $M\ej$, with $R_2$ calculated from equation~\eqref{R_2}. 
           The steeper part of each curve corresponds to $R_2=R\twomax$,
           obtained when cooling is neglected, i.e.~when $t\thm> t\inj$ in our model.
           If $t\thm=t\inj$, then $R_2=R\twothm$.
           All models have $\beta_1=0.9$, $\lambda_1=1$, $C'=\lambda_2$, and $T_1=T_2=4\times10^3\K$.
           \textit{Bottom}:~The ratio $R_2/R_1$ plotted against the original radius $R_1$, 
           for the case $M\ej=0$,
           assuming $\beta_1=0.9$, $\lambda_1=1$, $C'=\lambda_2$, $T_1=4\times10^3\K$, and $t\inj=10\yr$.
           Wherever $t\thm>t\inj$, $R_2/R_1$ is independent of $R_1$ and the curve is flat.
           \label{fig:Rratio}
          }
  \vspace{-0.3cm}
\end{figure}

We postulate that with cooling, either the star expands until it  reaches the radius calculated in Sec.~\ref{sec:expansion_nocooling} or 
until the  energy loss and injection rates  balance.
Thus,  
\begin{equation}
  \label{R_2}
  R_2=\min(R\twothm,R\twomax),
\end{equation}
where $R\twomax$ is the value of $R_2$ calculated in Sec.~\ref{sec:expansion_nocooling}, 
and $R\twothm$ is estimated by equating the mean rates of energy injection  and loss due to radiative cooling and winds, namely
\begin{equation}
  \label{Lbalance}
  \beta_1\frac{GM_1 M\eone}{\lambda_1R_1t\inj}\sim 4\pi\sigma(R\twothm^2T_2^4-R_1^2T_1^4) +\frac{C'GM_2 M\ej}{\lambda_2R\twothm t\inj},
\end{equation}
where $t\inj$ is the  energy injection time  during CE1, (see Sec.~\ref{sec:timescales}); 
the left side is the  energy injection rate $E\inj/t\inj$;
the first term on the right  is the luminosity change   $\rmD L$;
$T_1$ and $T_2$ are the effective temperatures of the star before and after CE1;
and the last term is the rate of energy transfer to ejecta, $E\ej/t\inj$.
The quantity $R_2$ transitions from $R\twothm$ to $R\twomax$ when $t\inj$ falls below the 
cooling time $t\thm=(E\inj-E\ej)/\rmD L$.
Equation~\eqref{Lbalance} reduces to a cubic  in $R\twothm$,
\begin{equation}
  \label{cubic}
  R\twothm^3 +bR\twothm +c \sim0,
\end{equation}
where
\[
  b=-\left[\frac{\beta_1 A M_1 M\eone}{\lambda_1R_1} +R_1^2\left(\frac{T_1}{T_2}\right)^4\right], \qquad c=\frac{C' A M_2 M\ej}{\lambda_2},
\]
with $A\equiv G/4\pi\sigma T_2^4t\inj$.
The relevant solution is
\begin{equation}
  R\twothm= 2\sqrt{-\frac{b}{3}}\cos\left[\frac{1}{3}\arccos\left(\frac{3c}{2b}\sqrt{-\frac{3}{b}}\right)\right].
\end{equation}
If $E\ej=0$, then $c=0$ and $R\twothm=\sqrt{-b}$.

In the top panel of Fig.~\ref{fig:Rratio}, we plot $R_2/R_1$ against $M\ej$ for four different models.
All models assume $\beta_1=0.9$, $\lambda_1=1$, and $C'=\lambda_2$.
The solution can transition from $R\twomax/R_1$ (steeper portion of the curves) 
to $R\twothm/R_1$ (flatter portion) when $M\ej$ drops below some critical value.
We consider our two example primary stars from Sec.~\ref{sec:expansion_nocooling}.
However, now the solution also depends on $R_1$, which we set to $60\Rsun$ ($180\Rsun$) for the $1\Msun$ ($3\Msun$) star,
and $\lambda_2$, which we set to $1$ ($3$), 
so that the ratio $\lambda_2/\lambda_1=1$ ($3$) is preserved from those examples.
In addition, we illustrate two different values of $t\inj$: $10\yr$ and $100\yr$;
these choices are motivated in Sec.~\ref{sec:timescales}.
Finally, we adopt $T_1=T_2=4\times10^3\K$ for each curve.
We see that $R_2/R_1\sim1.5$--$5$ if $M\ej\lesssim0.2\Msun$.

In the bottom panel, we adopt $M\ej=0$, i.e.~zero mass loss during CE1,
and $t\inj=10\yr$, and plot $R_2/R_1$ against $R_1$, 
changing one parameter value at a time in each of the models shown.
The solution changes from $R\twomax/R_1$ (flat) to $R\twothm/R_1$ at some critical value of $R_1$.
As in the top panel, all models assume $\beta_1=0.9$, $\lambda_1=1$, and $C'=\lambda_2$.
Reducing $T_2$ reduces $\rmD L\rad$, which can increase  $R_2/R_1$, 
so our choice of $T_2=T_1$ for most of the curves is conservative.  
The black (red) curve in the top and bottom panels corresponds to the same overall model, 
sliced through $R_1=60\Rsun$ ($R_1=180\Rsun$) in the top panel and $M\ej=0$ in the bottom.

\subsection{Injection time scale}
\label{sec:timescales}

For the cases of Sec.~\ref{sec:mass},  we estimate the duration  of energy injection into the envelope, $t\inj$, 
motivating the  values of Sec.~\ref{sec:expansion_withcooling}.
In case~(i) the tidally disrupted planet is accreted onto the WD, liberating its  orbital energy. 
If accretion sustains \citep{Guidarelli+19}, the  Eddington rate is \citep[e.g.][]{Chamandy+18}
$\dot{M}\Edd\sim 2.7\times10^{-5}\Msun\yr^{-1} \left(R\w/0.013\Rsun\right)$,
and
\begin{equation}
  t\inj\sim \frac{m_1}{\dot{M}\Edd}
    \sim 37\yr\left(\frac{m_1}{10^{-3}\Msun}\right)\left(\frac{R\w}{0.013\Rsun}\right)^{-1}.
\end{equation}
But angular momentum redistribution during tidal disruption 
could swiftly drive some material to the core (Guidarelli et al., in preparation).
For case~(i), we thus crudely estimate $1\lesssim t\inj \lesssim 100\yr$.
For case~(ii), orbital energy is injected only down to the tidal disruption separation, and $t\inj$ is smaller.

The envelope responds to energy injection on its sound-crossing time, 
of order of days to months.
If the energy is deposited at the base of the convective zone, 
then convection  transfers it to the envelope on a convective time, a few sound-crossing times.
If energy is deposited  in a deeper radiative zone, 
this   delays the energy to the envelope by of order a photon diffusion time $\sim10\yr$ using \citet{Chamandy+19b}.
The envelope would continue to adjust  as this energy is transferred.
These processes might stall the envelope response  but are unlikely to change its duration.
We have thus assumed in Sec.~\ref{sec:expansion_withcooling},
that the envelope responds on a  time $\sim t\inj$.

\vspace{-0.4cm}
\section{Conclusions}
\label{sec:conclusions}
Planets incurring successive CE events can combine to eject a stellar envelope.
A planet interior to WD~1856~b could have partially unbound and expanded the envelope, 
which engulfed WD~1856~b and caused CE2, 
which completed the ejection and left the planet stabilized in its observed state.
We find that $0.2\lesssim m_1/\MJ\lesssim 30$ 
emerges as a plausible range for the mass of the planet/brown dwarf involved in CE1. 

We estimated the stellar radius increase  during CE1 using energy arguments, 
allowing for mass loss and radiative cooling, 
finding the ratio of final to initial radii  $1.5\lesssim R_2/R_1\lesssim 10$.
CE1 can thus plausibly initiate CE2.
However, the envelope cannot expand if the ejecta carries away too much energy and offers too little drag
on the secondary to compensate.

Since the orbit of planet~2 stabilizes around the  time of CE2, 
our model predicts a more uniform distribution in WD ages than does the ZLK-tidal friction mechanism, 
which requires much longer time scales to operate.
Our two-planet scenario could be generalized to more planets
with only the final event ejecting the envelope and avoiding disruption.
Massive planets are  more likely to eject what remains 
so the last surviving planet would likely be relatively massive. 

Evidence for previous mergers of planets with the stellar core in our scenario may be chemical enrichment of the WD, 
or accretion discs \citep[e.g.][]{Koester+09, Girven+12, Doyle+19, Veras+Heng20},
if sufficiently young, 
or high WD magnetic fields if the fields are acquired by accretion  of a tidally disrupted companion \citep{Nordhaus+11}.
Further work on the angular momentum distribution of WDs may be of interest  in this context, 
as accretion can spin up the WD whilst magnetic breaking can spin it down.

Our results are broadly consistent with \citet{Siess+Livio99a,Siess+Livio99b},
who compute detailed spherically symmetric models involving a $3\Msun$ AGB star or $\sim1\Msun$ RGB star
undergoing accretion of planetary or brown dwarf material deep inside the envelope.
A roughly two-fold radial expansion of the star over $\sim100\yr$ is fairly typical for their models,
and in some cases accretion increases nuclear burning, 
resulting in a roughly four-fold expansion over $\sim10\yr$.
That the primary's envelope  would expand significantly during a CE interaction 
with a planet is further supported by \citet{Staff+16b}, 
who simulated CE interactions between a $10\MJ$ planet and a $3.5\Msun$ RGB star or $3.05\Msun$ AGB star,
finding that the envelope expands by $\sim40\%$ or $\sim20\%$, respectively.
Higher resolution simulations that can follow the inspiral down to smaller separations,
and further release of orbital energy, would be desirable.

\vspace{-0.3cm}
\section*{Acknowledgements}
We acknowledge  A. Frank for discussions, the referee for key suggestions, N. Soker for helpful comments, and
US DOE grants DE-SC0001063, DE-SC0020432, DE-SC0020103; NSF grants AST-2009713, AST-1515648, AST-1813298, PHY-2020249; 
STSCI grant HST-AR-12832.01-A, HST-AR-15044.

\vspace{-0.3cm}
\section*{Data Availability}
Observational data  are taken from \citet{Vanderburg+20}.

\vspace{-0.2cm}
\footnotesize{
\bibliographystyle{mnras}
\bibliography{refs}
}
\end{document}